\documentclass[12pt,preprint]{aastex}

\begin{document}

\title{On the properties of fractal cloud complexes}

\author{N\'estor S\'anchez,\altaffilmark{1,2}
        Emilio J. Alfaro,\altaffilmark{1} and 
        Enrique P\'erez\altaffilmark{1}}
\altaffiltext{1}{Instituto de Astrof\'{\i}sica de Andaluc\'{\i}a,
                 CSIC, Apdo. 3004, E-18080, Granada, Spain;
                 nestor@iaa.es, emilio@iaa.es, eperez@iaa.es.}
\altaffiltext{2}{Departamento de F\'{\i}sica, Universidad del
                 Zulia, Maracaibo, Venezuela.}

\begin{abstract}
We study the physical properties derived from interstellar cloud
complexes having a fractal structure. We first generate fractal
clouds with a given fractal dimension and associate each clump
with a maximum in the resulting density field. Then, we discuss
the effect that different criteria for clump selection has on the
derived global properties. We calculate the masses, sizes and
average densities of the clumps as a function of the fractal
dimension ($D_f$) and the fraction of the total mass in the form
of clumps ($\epsilon$). In general, clump mass does not fulfill
a simple power law with size of the type $M_{cl} \varpropto
R_{cl}^{\gamma}$, instead the power changes, from $\gamma \simeq
3$ at small sizes to $\gamma < 3$ at larger sizes. The number
of clumps per logarithmic mass interval can be fitted to a power
law $N_{cl} \varpropto M_{cl}^{-\alpha_M}$ in the range of
relatively large masses, and the corresponding size distribution
is $N_{cl} \varpropto R_{cl}^{-\alpha_R}$ at large sizes. When
all the mass is forming clumps ($\epsilon = 1$) we obtain that
as $D_f$ increases from $2$ to $3$ $\alpha_M$ increases from $\sim 0.3$
to $\sim 0.6$ and $\alpha_R$ increases from $\sim 1.0$ to $\sim 2.1$.
Comparison with observations suggests that $D_f \simeq 2.6$ is roughly
consistent with the average properties of the ISM. On the other hand,
as the fraction of mass in clumps decreases ($\epsilon < 1$) $\alpha_M$
increases and $\alpha_R$ decreases. When only $\sim 10\%$ of the complex
mass is in the form of dense clumps we obtain $\alpha_M \simeq 1.2$ for
$D_f=2.6$ (not very different from the Salpeter value $1.35$), suggesting
this a likely link between the stellar initial mass function and the
internal structure of molecular cloud complexes.
\end{abstract}


\section{Introduction}

The fact that the interstellar medium (ISM) has a hierarchical and
self-similar structure when observed with sufficiently high dynamic
range, is interpreted as evidence of an underlying fractal structure
\citep{sca90}. The boundaries of the projected images of interstellar
clouds are irregular curves whose fractal dimension is around $\sim
1.3$ \citep[e.g.][]{fal91,lee04}. Apparently this is a universal
result which does not depend on whether tracers of atomic, molecular
or dust components are used, whether clouds are selfgravitating or not,
etc \citep{wil00}. The fractal dimension of the projected boundaries
is usually associated with a three-dimensional fractal dimension
$D_f \simeq 2.3$ for the ISM \citep[e.g.][]{bee92}. It has been
argued that a fractal ISM with $D_f \simeq 2.3$ could account for
the observed mass and size distributions of the interstellar
clouds \citep{elm96}, as well as for the intercloud medium properties
\citep{elm97a}, and even for the stellar initial mass function
\citep{elm97b,elm99}.

In a previous paper \citep{san05} we studied the effect
that the projection of clouds has on the estimation of the fractal
dimension, and we concluded that a value around $\sim 1.3$ for the
projected boundaries is more consistent with three-dimensional clouds
having $D_f \sim 2.6 \pm 0.1$. The application of $\Delta$-variance 
techniques to Polaris Flare cloud by \citet{stu98} yielded a fractal
dimension for the cloud surfaces $\simeq 2.6$ \citep[see also][]{ben01}.
\citet{elm02} simulated fractal brownian motion clouds with average
fractal dimension $\sim 2.75$ and obtained properties in gross
agreement with observations. A fractal medium with a relatively
high $D_f$ value can reproduce observations of HII regions
surrounding stars \citep{woo05}. \citet{hen86,hen91} used a
gravitationally driven turbulence model to study the properties
of giant molecular clouds suggesting that a dimension $\simeq
2.7$ could be necessary to explain the observed properties. Also
\citet{fle96} analyzed the properties of the turbulent,
non-self-gravitating, neutral component of the ISM by using
a model of compressible turbulence, concluding that the
compression parameter that better reproduces observations is
such that $D_f \simeq 2.5$.

It is important to quantify the degree of complexity (through,
for example, the fractal dimension) as a first step towards
understanding the physical mechanisms responsible for
structuring the ISM. In models of fractally homogeneous
turbulence the fractal dimension of fully developed
turbulence is $2.5 < D_f < 2.75$ \citep{hen82}. On the
other hand, turbulent diffusion in a incompressible medium
generates structures with $D_f \sim 2.3$ for a Kolmogorov spectrum
\citep{men90}. It is not obvious, however, that the energy can
cascade down without any dissipation or injection, and deviation
from a Kolmogorov spectrum is, in general, expected in the ISM
\citep{bru02a,bru02b}. The ISM is a highly compressible and turbulent
medium and its fractal dimension depends, among other factors,
on the degree of compressibility \citep{fle96} or on the
Mach number \citep{pad04}. Also, self-gravity could by itself
explain many observed properties \citep{dev96}, although this
fact does not imply that turbulence is not an important factor
in the ISM.

Here we are interested in understanding the relationship between 
the physical properties of the interstellar clouds and their
fractal structure, and in verifying whether observed properties
are in agreement with relatively high fractal dimension values
\citep[as suggested in][]{san05}. In other words, we investigate
what physical properties are intimately connected to the cloud
``geometry" and whether this geometry, no matter how it was
originated, defines its next evolutionary stage. Our approach in
this work is to calculate and analyze the properties resulting from
a hierarchical structure with a known and perfectly defined fractal
dimension, no mattering the physical processes behind this structure.
In order to do this, we simulate interstellar clouds by using a simple
algorithm explained in section~\ref{sec_fractals}, which generates a
distribution of points with a very well defined fractal dimension.
In section~\ref{sec_pdf} we address the density fields resulting
from the generated fractals and in section~\ref{sec_clump} we discuss
the different ways to construct ``clumps" from these density fields.
Section~\ref{sec_results} is devoted to study the properties (masses,
densities, sizes, and mass and size distributions) of the derived clumps
and to compare these results with observations. Finally, the main
conclusions are summarized in section~\ref{sec_conclusion}.

\section{Simulated fractal clouds}
\label{sec_fractals}

To generate fractal clouds we have used the same procedure described
in \citet{san05}. Within a sphere of radius $R_f$ we randomly place
the centers of $N$ spheres of radius $R_f / L$ with $L > 1$. In each
sphere we again place the centers of $N$ smaller spheres with radius
$R_f / L^2$, and so on, up to a given level $H$ of hierarchy. At the
end of this procedure there are $N^H$ points distributed in the space
with a fractal dimension given by $D_f = \log(N)/\log(L)$. This kind
of fractals mimics in some way the hierarchical fragmentation process
occurring in molecular cloud complexes. We have used $N = 3$ fragments
through $H = 9$ levels of hierarchy with the fractal dimension
ranging from $2.0$ to $3.0$. To reduce possible random variations
all the properties shown are the result of calculating the average
of $10$ different realizations (random fractals). In order to prevent
the appearance of a multifractal behavior it is necessary that
spheres do not superpose when generating the fractal cloud, and this
requirement prevents the algorithm from generating random fractals
with $D_f > 2.6$. Fractals with $D_f = 3$ have been obtained by
distributing randomly $N^H$ points in the available volume, and
in this case we have calculated the average properties of $50$
different realizations.

The properties resulting from doing an {\it ideal random sampling
throughout the hierarchy} in these kind of fractals have been
discussed by \citet{elm97a}. Let us call $R_f$ and $M_f$ the radius
and mass (number of points) of the fractal, respectively. The average
density of the whole fractal is $\rho_f = M_f/(4/3)\pi R_f^3$. In the
$i$-th level ($i = 0,1,...,H$) there are $n_i = N^i$ fragments with
radii $R_i/R_f = N^{-i/D_f}$, masses $M_i/M_f = N^{-i}$, and average
densities $\rho_i/\rho_f = N^{i(3/D_f-1)}$. Then, the mass increases
with the radius as $M_i/M_f = (R_i/R_f)^{D_f}$ and the density varies
as $\rho_i/\rho_f = (R_i/R_f)^{-(3-D_f)}$. The number of fragments
with a given mass value is $n(M) = (M/M_f)^{-1}$, so that the mass
distribution function in equal intervals of $M$ is $dn(M)/dM \sim
M^{-2}$, independent of the fractal dimension $D_f$. This means that
the mass distribution function per logarithmic mass interval is
$dn(M)/d\log M \sim M^{-1}$. On the other hand, the number of
fragments depends on the radius as $n(R) =  (R/R_f)^{-D_f}$ and
the size distribution function per logarithmic interval will be
$dn(R)/d\log R \sim R^{-D_f}$. Therefore, the theoretically predicted
mass and size distributions from random sampling in a fractal cloud
are power laws of the form $dn(M)/d\log M \sim M^{-\alpha_M}$ and
$dn(R)/d\log R \sim R^{-\alpha_R}$ with indices (i.e., slopes in a
log-log plot) given by $\alpha_{M} = 1$ and $\alpha_{R} = D_f$,
respectively. However, in the hierarchical structure the smaller
fragments are not isolated objects but they are included into the
big ones, and if double counting is avoided then the indices
$\alpha_{M}$ and $\alpha_{R}$ become lesser \citep{elm97a}. Even
though we do not take this effect into account, smaller fragments
can be blended into big ones in real clouds. This blending can be
``real" (random cloud-cloud coalescence) or ``artificial" (due to
resolution limitations), but the effect should be the same: to
decrease the expected values of $\alpha_{M}$ and $\alpha_{R}$.
Therefore, it is not clear what values should we measure when
observing fractal complexes as the ones described above, and this
is the main goal of this work.

\section{Probability density function}
\label{sec_pdf}

\begin{figure}
\plotone{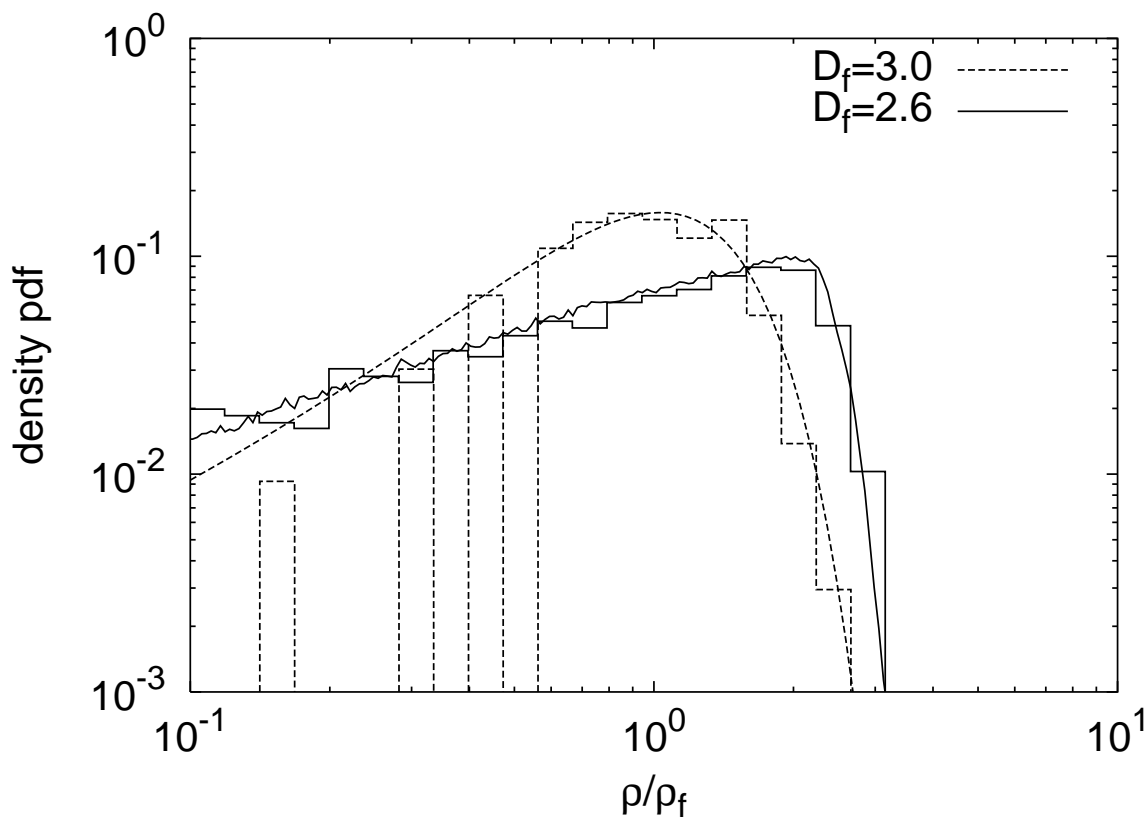}
\caption{The density pdf calculated in a direct way by using cells
         with radius $R_c = R_f/10$ for the fractals with dimensions
         $D_f=3$ (dashed line histogram) and $D_f=2.6$ (solid line
         histogram). The dashed curve is a gaussian fit for $D_f
         =3$ and the solid curve shows the density pdf calculated for
         $D_f=2.6$ but using a gaussian kernel with $\sigma=R_c$ (see
         text).}
\label{fig_pdf}
\end{figure}

Before proceeding with the calculation of the cloud properties,
we have to estimate the one-point probability density function
(hereinafter pdf) of the density for the fractal distribution
of points. This is an important issue because we will need the
density pdf to compute clump masses and radii in the following sections.
The most direct way to do this is to place a grid of cells on the fractal,
then calculate the density in each cell and plot a histogram with
the number of cells in a certain range of density. An alternative
way that guarantees a good sampling is to place randomly a high enough
number of cells within the fractal. The density in each sampling
is then the number of points inside the cell divided by the volume.
As critical requirement for a suitable estimation of the pdf, the
cell size ($R_c$) has to be big enough to produce a large number
of density values, i.e., it has to be much bigger than the minimal
distance between two points in the fractal. Figure~\ref{fig_pdf}
shows the density pdf (per logarithmic interval of density) resulting
from using spherical cells with radius $R_c = R_f/10$ for the fractals
with dimensions $D_f=3$ and $D_f=2.6$. For the case $D_f=3$ (very
homogeneous distribution of points) we see that, as expected, the
density pdf is very similar to a gaussian with the maximum around
$\rho/\rho_f = 1$. For comparison we also show, with a dashed curve,
a gaussian distribution fitted to the $D_f=3$ histogram. For the case
$D_f=2.6$ the resulting pdf departs from a gaussian distribution. The
behavior resembles a power law at low densities (with slope $\sim
0.65$) with a sharp cut-off after a maximum. The expected density
value, when a whole structure with size $R_c$ is sampled, is
$\rho_c/\rho_f = (R_c/R_f)^{-(3-D_f)}$, which for $D_f=2.6$ means
$\rho_c/\rho_f \simeq 2.5$. The pdf cut-off is close to this value,
although higher values can be seen due to the presence of random
denser regions. However, much higher density values cannot be detected
by using this cell size, it is necessary to use smaller cells to be
able to sample efficiently denser structures belonging to lower
levels. The average density of the (upper) level where the cell
of size $R_c$ is embedded is given by $\rho_{c-1}/\rho_f =
(N^{1/D_f} R_c/R_f)^{-(3-D_f)}$, which gives $\simeq 2.12$ for $D_f=2.6$.
This is indeed the value around which the maximum of the distribution
is found. Smaller density values can be obtained as the (fixed
volume) cell samples less massive structures in lower levels. It
has to be mentioned that cells of different sizes tend to produce
distributions with different positions for the maximum (higher
values as $R_c$ decreases), but the distributions always follow
roughly a power law at low densities with a cut-off after the
maximum.

\begin{figure}
\plotone{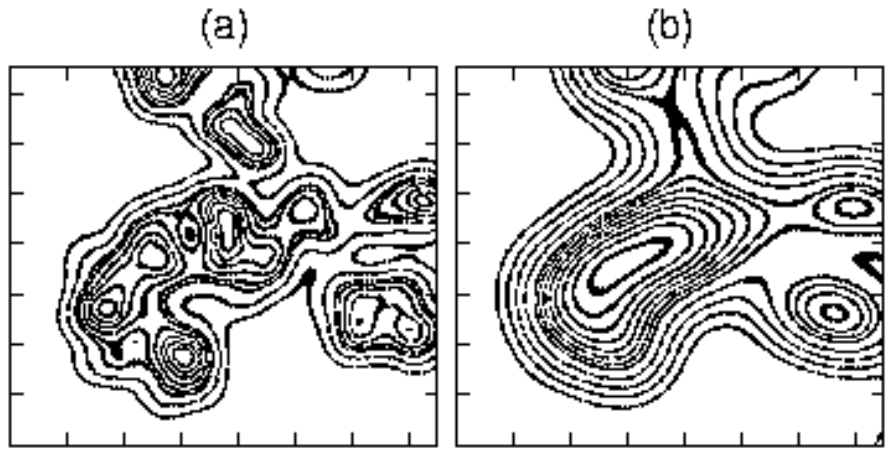}
\caption{A slice along the $z=0$ plane showing density contours
         of part of a fractal with $D_f=2.6$, calculated by using
         equation~(\ref{ecu_kernel}) with (a) $\sigma=\sigma_{opt}$
         and (b) $\sigma=2\sigma_{opt}$.}
\label{fig_mapa}
\end{figure}

The pdf estimated in this direct way exhibits a discrete distribution
because there can only be an entire number of points inside the cell.
In order to turn it into a continuous distribution we use a kernel
function to convolve the data and calculate the density through the
whole available volume. We have adopted a gaussian kernel (normalized
to have an integral of 1); thus, the density at the position
$(x_i,y_i,z_i)$ is \citep{sil86}
\begin{eqnarray}
\label{ecu_kernel}
f(x_i,y_i,z_i) = \frac{1}{N^H \sigma^3 (2\pi)^{3/2}}
\sum_{j=1}^{N^H} \exp \left\{ -\frac{1}{2\sigma^2} \left[
(x_i-x_j)^2 + (y_i-y_j)^2 + (z_i-z_j)^2 \right] \right\} \ \ \ ,
\end{eqnarray}
where $\sigma$ is the ``smoothing" parameter, which determines the
volume used to estimate the average density in each position (roughly
equivalent to the former cell size). As an example,
Figure~\ref{fig_pdf} shows with solid line the pdf calculated for
$D_f=2.6$ by using equation~(\ref{ecu_kernel}) with $\sigma=R_c$. As
before, the pdf was constructed by calculating the local density
in randomly chosen positions and plotting the number of density
values per logarithmic interval of density. The result is very
similar to that from the direct method except that now
we have a continuous distribution. Again, the exact functional
form will depend on the adopted values for $\sigma$. Large $\sigma$
values (of the order of the fractal size) lead to a too homogeneous
distribution of density values, whereas small values (much smaller
than the minimal distance between two points) tend to produce very
narrow density peaks. To avoid both undesired situations and to be
as objective as possible we always have used the $\sigma$ values
that maximize the likelihood \citep{sil86}, that we call
$\sigma_{opt}$ herein. Figure~\ref{fig_mapa}
is an example of the appearance of the density field
resulting from applying equation~(\ref{ecu_kernel}) to a
fractal with $D_f=2.6$. Each map is a slice along the $z=0$ plane
showing density contours of a part of the three-dimensional
density field. The left-side map (a) was generated with
$\sigma=\sigma_{opt}$, whereas the right-side map (b) is exactly
the same region but generated with $\sigma=2\sigma_{opt}$. Clearly
we see that as $\sigma$ increases some density peaks tend to blend
with other peaks forming smoother regions. Several tests showed
that the properties studied in this work do not depend (within the
error bars) on the exact value of $\sigma$ as long as this value
is of the order of $\sigma_{opt}$. This value can be
different from fractal to fractal but always was of order
$\sim 10^{-2} R_f$.

\begin{figure}
\plotone{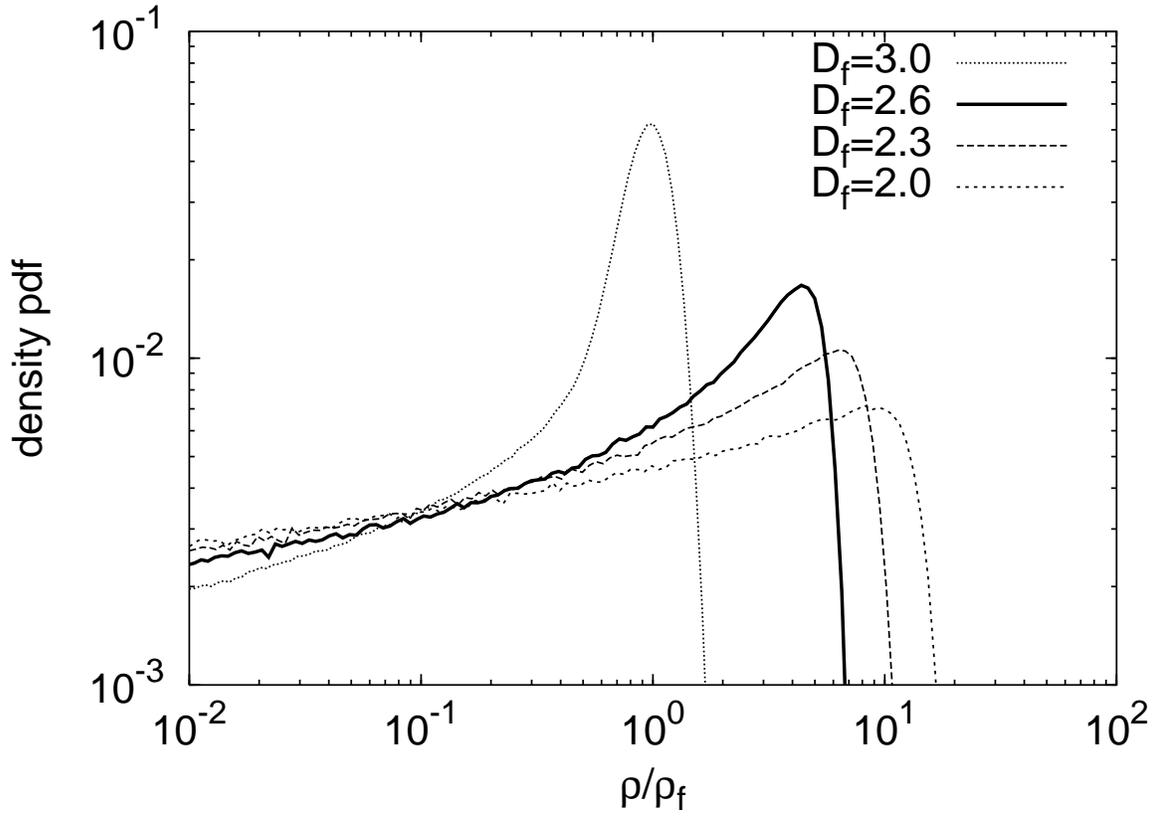}
\caption{The density pdf calculated by using a gaussian kernel
         with $\sigma = \sigma_{opt}$, for different values of
         fractal dimension: $D_f=3$ (dotted line), $D_f=2.6$
         (solid line), $D_f=2.3$ (long dashed line), and $D_f=
         2.0$ (short dashed line).}
\label{fig_kernel}
\end{figure}

In Figure~\ref{fig_kernel}
we show the normalized density pdf calculated (with $\sigma =
\sigma_{opt}$) for the fractals with dimensions $D_f=3$, $D_f=2.6$,
$D_f=2.3$, and $D_f=2.0$. The long tails at low densities come
from the sampling in or close to the edge and/or the voids
existing in the fractal clouds. The maximum of the distribution
moves toward higher density values as $D_f$ decreases according
to the relation $\rho_{max} \varpropto R^{-(3-D_f)}$ for $R
\sim constant$, from $\rho_{max} / \rho_f = 1$ for $D_f=3$ to
$\rho_{max} / \rho_f \simeq 10$ for $D_f=2.0$. There is always
a turn over in the pdf, decaying rapidly at densities just after
the maximum.

The density pdf is a fundamental statistical quantity characterizing
the structure of a given medium. Despite the pdf being a one-point
statistic containing no spatial information, the knowledge of
this function is a fundamental step prior to the analysis of the
star formation process and the stellar initial mass function
\citep[e.g.][]{pad02,sca98}. The problem is that
the density pdf cannot be directly measured in the ISM,
instead the column density pdf is measured. \citet{ost01}
used three-dimensional numerical simulations of isothermal
magnetohydrodynamic turbulence to compare the resulting density
field with the column densities, finding that both distributions had
very similar shapes (approximately lognormal). Also \citet{fis04}
showed that lognormal density distributions produce lognormal column
density pdf's. However, column density observations cannot always
be fitted to lognormal distributions and can show extended tails
\citep{bli97} or simply a power law behavior \citep{hei05}.

The suggestion that the density pdf in the ISM should be consistent
with a lognormal distribution comes from the central limit theorem
applied to a multiplicative hierarchical density field \citep{vaz94}.
Several authors have found that the density pdf of turbulent gas should
exhibit a roughly lognormal distribution under a wide range of
physical conditions \citep[][and references there in]{bal04,elm04}.
For a polytropic gas (where $P \varpropto \rho^n$) the shape
of the pdf depends on the effective polytropic index $n$.
If the gas is isothermal ($n=1$) the pdf is lognormal \citep{vaz94}
but extended tails can develop at high or low densities if $n \neq
1$ \citep{pas98,sca98,ost01}, and the departure from a lognormal
is larger at higher Mach numbers \citep{ber05}. The one-dimensional
simulations of highly compressible turbulence of \citet{pas98} showed
that a power law tail develops at relatively high densities if
$n<1$ and at low densities if $n>1$. Their results for $n>1$ and high
Mach numbers resemble the pdf's we showed in Figure~\ref{fig_kernel}.

The situation becomes more complicated if we try to understand
the dependence of the density pdf on the physical mechanisms acting
in the turbulent medium. \citet{li03} found that shocks sweep up
the gas producing low density regions and pdf's skewed to low
densities, but self-gravity tends to collect the gas producing pdf's
with positive skewness. The presence of a magnetic field has,
however, very little effect on the density pdf \citep{pas03}.
The large-scale simulations performed by \citet{wad01} showed a
roughly constant pdf shape (lognormal at high densities and normal
at low densities) regardless of whether star formation is considered
or not in the simulation. \citet{sly05} arrive to very different
conclusions, finding that the pdf does depend sensitively on the
simulated physics. Their pdf is consistent with a lognormal function
only for the runs where neither supernova feedback nor self-gravity
are implemented. With the addition of self-gravity the pdf approaches
a power law at high densities, but it becomes markedly bimodal when
supernova feedback is included (with or without self-gravity).

\section{Clump definition}
\label{sec_clump}

Starting from the three-dimensional density field calculated
by applying equation~(\ref{ecu_kernel}) to the generated fractals,
we want to study the properties of the resulting ``density
condensations". We will call these regions ``clumps" to
differentiate it from the whole fractal structure, which we could
associate with a giant molecular cloud. Actually this denomination is
arbitrary, firstly because there are no precise definitions of terms
like ``cloud", ``clump", or ``core" \citep{lar03}, and secondly
because here we can not speak about an absolute spatial scale but
about self-similar structures inside a region of size $R_f$.
Different schemes can be adopted in defining a clump. For example,
the algorithm called GAUSSCLUMPS \citep{stu90} uses a least squares 
fitting procedure to decompose iteratively the emission in one or
more gaussian clumps, instead CLUMPFIND \citep{wil94} associates
each local emission peak and the neighboring pixels with only one
clump (similar to the usual eye-inspection procedure). In both cases,
the implicit assumption that the radial velocity coordinate can be
replaced by the radial distance is made, but this assumption
is not necessarily always satisfied \citep{ost01}. In our case,
this problem does not exist because we are using three spatial
coordinates and density values, but we have to keep in mind that
comparison with observations may not be done directly.
For simplicity, we have chosen to associate each peak in density
with one clump (in the same way that CLUMPFIND does).
Once we have identified a peak we determine the mass of the clump
by integrating the density field over the neighboring region. In
order to proceed with the calculation we need to place a grid over
the volume occupied by the fractal cloud. The size of the cell was
chosen to equal de value of $\sigma_{opt}$, that is, to equal the
resolution with which the density field was generated. Since
$\sigma_{opt}/R_f$ was always of order of $10^{-2}$ or less,
there were always $\sim 10^6$ cells or more over the fractal.

An important aspect to consider when defining a clump is
the noise present in the data, whether it be observational or
numerically simulated. Usually the problem is approached by
defining a free parameter such that fluctuations of the order of
(or smaller than) this parameter are not considered as clumps.
For example, in the algorithm CLUMPFIND this parameter is the
step between two successive contours, in the simulations of
\citet{gam03} is their smoothing length, and in this work
it is the parameter $\sigma$ used in equation~(\ref{ecu_kernel})
to generate the density field. Another very important
point is to define the boundary of the
cloud, i.e., to define which neighboring region ``belongs"
to the clump and which not. Again the usual approach is to
introduce an additional parameter: the threshold density
(or emission). Thus, a common definition for clump in literature
is all the pixels around a maximum in density with
values greater than some predefined threshold value. The
algorithm we have constructed to identify clumps works in
the following manner. First, it associates the pixel with
highest density value to the first clump. Then it takes
the pixel whose density value is just below the highest, and
if this pixel is neighbor of the first one then it ``belongs"
to the first clump. If by the contrary this pixel is isolated
then it is labeled as the first pixel of the second clump.
All the pixels are explored successively in decreasing order
of density in the same way: if a pixel is contiguous to the
boundary of some existing clump then it belongs to this clump,
otherwise it belongs to a new clump. This procedure continues
until some threshold density is reached.

\begin{figure}
\plottwo{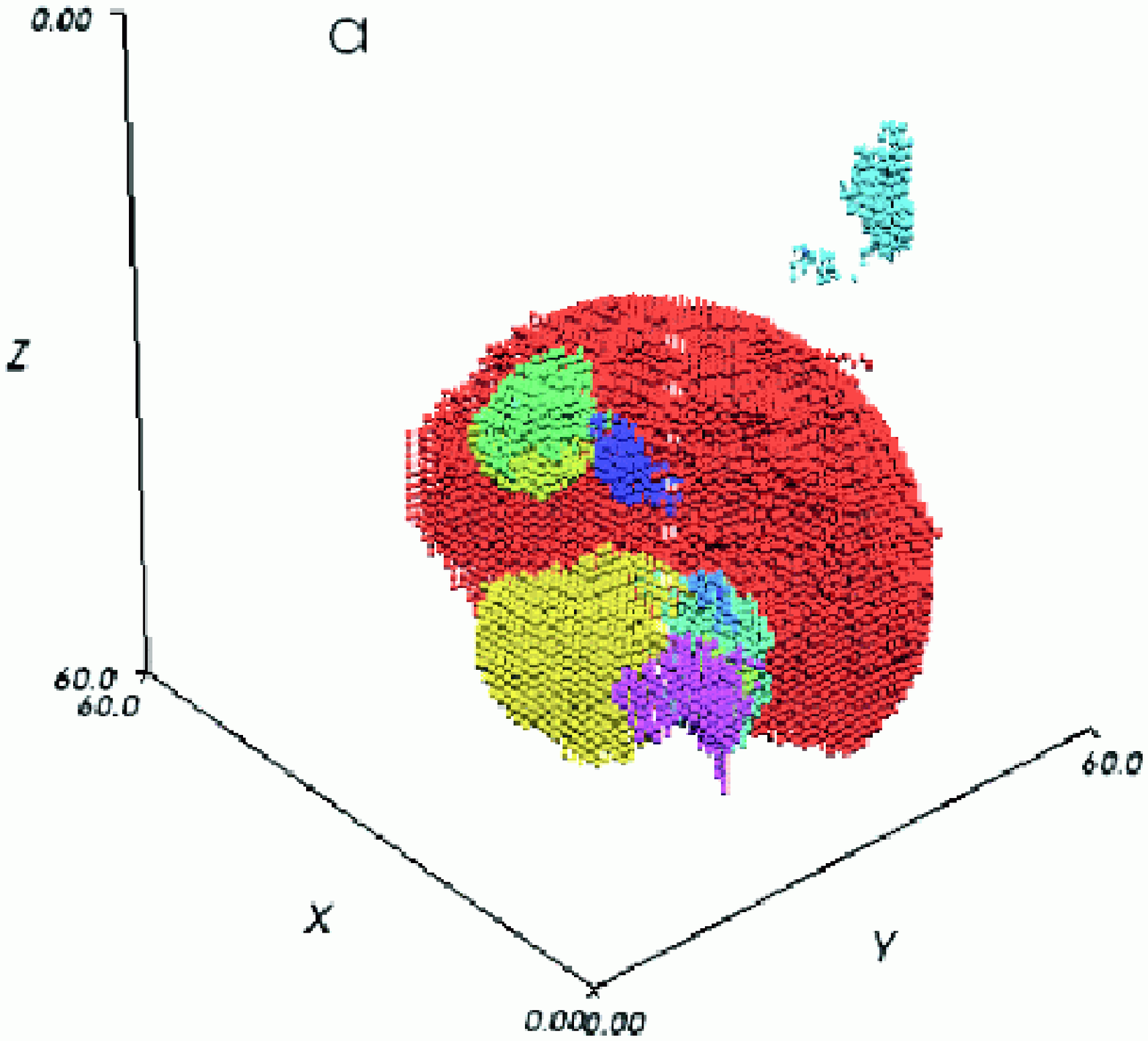}{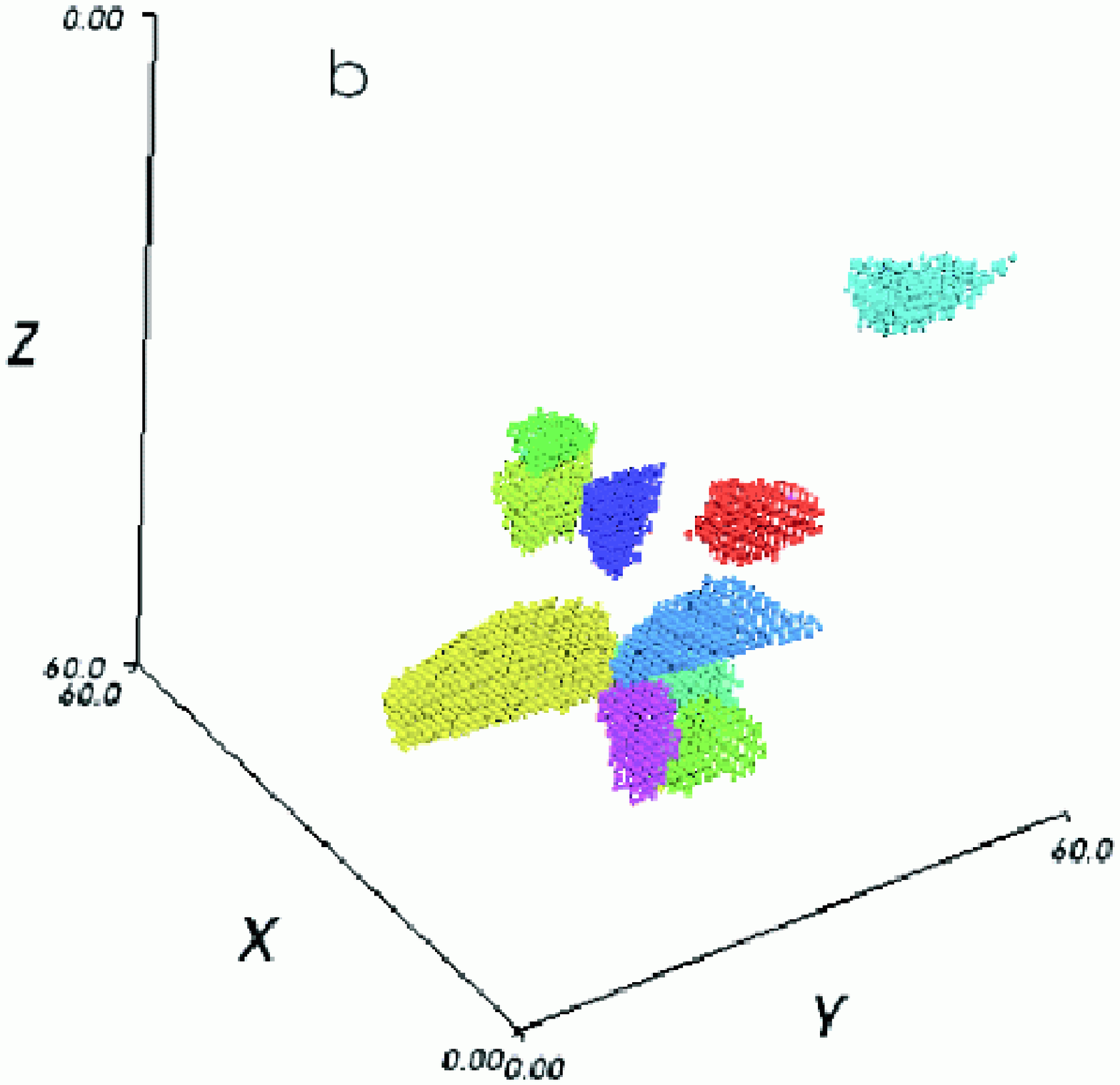}
\caption{Images of the clumps resulting (each one with a different
         color) from ten density peaks when $\rho_{th}=0$, for the
         first random fractal generated with $D_f=2.6$, using
         (a) the criterion ``potential" and (b) the criterion
         ``distance" (see text). The coordinates are the pixel
         coordinates.}
\label{fig_clumps100}
\end{figure}

\begin{figure}
\plotone{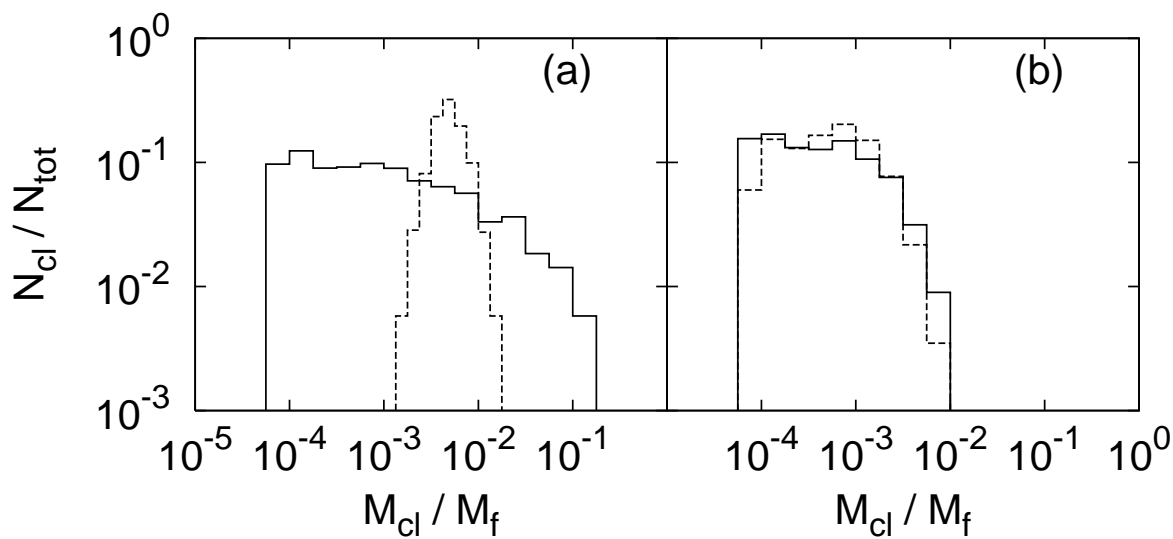}
\caption{The clump mass spectra for ten random fractals with
         $D_f=2.6$ when using the criterion of potential (solid
         line histograms) and the criterion of distance (dashed
         line histograms). (a) All the cloud gas is forming
         clumps and (b) only 10\% of the cloud gas is in the
         form of clumps.}
\label{fig_criterios}
\end{figure}

When this threshold is relatively low there can be an additional
problem: two (or more) peaks in density may be present in the
region above the threshold, and we have to ``decide" to which
of the available peaks the pixels that are contiguous to two or more
boundaries should be assigned to. The criterion normally used
is purely geometric: each ``ambiguous" pixel is assigned to the
clump with the closest peak (just like CLUMPFIND does). However,
we should ask ourselves whether different criteria could be used
to assign ``membership" in a more ``suitable" way. An alternative
criterion could be to assign each ambiguous pixel to the clump
with highest ``gravitational potential". We have tested the former
criterion by estimating the potential (at the moment of considering
the $i$-th pixel) as the current clump mass ($M_i$) divided by the
peak-pixel distance ($d_i$). Then, each ambiguous pixel is
attached to the clump with higher $M_i/d_i$ value rather than
to the clump with higher $1/d_i$ value (i.e., smaller distance).
If the threshold density is close to the maximum density values
for two neighbor clumps, then the clumps will be physically
separated and any criteria for ambiguous pixels will yield the
same result. As threshold density ($\rho_{th}$) decreases
the adjacent surface between two neighbor
clumps increases and therefore the effect
of different criteria becomes more important. As an example,
Figure~\ref{fig_clumps100}
shows ten of the resulting clumps when $\rho_{th}=0$ for
the first random fractal we simulated with $D_f=2.6$ and for
both criteria mentioned above. The criterion of distance tends
to produce more symmetrical (and similar in shape) clumps,
and in this case the clump volume distribution relates
directly with the peak-peak distance distribution between
neighbors, which is random for the fractal sets we are
generating. This degenerates into a clump mass distribution
narrower than for the criterion of potential, as can be seen
in Figure~\ref{fig_criterios}a
where the number of clumps per logarithmic mass interval
(normalized to the total number of clumps) $N_{cl}/N_{tot}$
has been plotted as a function of the clump mass in units
of the total mass ($M_{cl}/M_f$). Thus, the same cloud density
field and therefore the same number of clumps yields very
different mass distributions only changing the criterion with
which the low density pixels are distributed. The differences
are amplified simply because we are artificially creating
clumps from extremely low density material which typically
remains in the form of inter-clump gas. The clumps obtained
by distributing all the cloud mass in clumps are far from
looking similar to those obtained from observations: there
are no observed clouds formed only by clumps without inter-clump
material. We have verified that in more realistic situations
when $\rho_{th} > 0$, i.e. when an inter-clump density is defined,
clumps resulting from applying different criteria and therefore
their global properties (such as mass and size distributions)
gradually begin to converge as $\rho_{th}$ increases. In fact,
when only 10\% of the cloud mass is forming clumps both criteria
for selection of clumps yield very similar mass distributions
(see Figure~\ref{fig_criterios}b). In other words, when we only
see the dense ``cores" the method of distributing low density
pixels does not play any important role.
In section~\ref{sec_threshold} we analyze the effect of the
threshold density on the global clump properties in order to
understand how these properties are modified in clouds observed
with different degrees of sensitivity and contrast.

We have tested several different criteria for clump selection but
the global results always are similar to one or another of the two
criteria above, depending on whether peak-pixel distance or
clump mass is the dominant factor in the criterion applied.
Here we have chosen to discuss in detail the properties
resulting when using the (physically inspired) criterion of
the ``potential". When considering the clump mass for deciding
about ambiguous pixels this criterion takes into account in
some way the underlying density pdf, which depends on the cloud
fractal dimension (section~\ref{sec_pdf}). The properties we
derived by using the criterion of distance map the random
distribution of distances between neighbor density peaks
rather than of intrinsic density structure. However, the
sensitivity of the cloud properties, both simulated and observed,
to different clump selection criteria is an important problem
that must be addressed in future works.

\section{Properties of fractal clouds}
\label{sec_results}

We have defined each clump as all the mass enclosed from the density
peak to a threshold density. For convenience, we have chosen the
simplest way to define this threshold: a constant value such that
a given ratio, $\epsilon$, of the total mass in clumps to the total
cloud mass is obtained. If $\epsilon = 1$ the number of clumps
formed will equal the number of density peaks, which is given
by the three-dimensional density field (i.e., by $D_f$) and it does not
depend on the clump selection criterion. Low $D_f$ values imply more
fragmented clouds (higher number of clumps). But additionally at
low $\epsilon$ values some peaks could be below the threshold density
and then there could be a lesser number of clumps. Table~\ref{tablaclumps}
shows the average number of clumps (per simulation) obtained for the
results we discuss in this work. We first discuss the results for
$\epsilon = 1$ and in section~\ref{sec_threshold} we analyze the 
effect of changing the threshold density.
\begin{deluxetable}{ccccc}
\tablewidth{0pt}
\tablecaption{Average number of clumps formed
per fractal cloud\label{tablaclumps}}
\tablehead{
\colhead{$D_f$} &
\colhead{$\epsilon=1$} &
\colhead{$\epsilon=0.5$} &
\colhead{$\epsilon=0.25$} &
\colhead{$\epsilon=0.1$}
}
\startdata
2.0 & 764.6 & 764.6 & 737.6 & 556.8 \\
2.1 & 620.6 & 620.6 & 592.7 & 469.3 \\
2.2 & 531.1 & 531.1 & 514.6 & 410.2 \\
2.3 & 427.5 & 427.2 & 412.5 & 316.8 \\
2.4 & 343.9 & 343.4 & 328.7 & 244.2 \\
2.5 & 262.1 & 262.0 & 243.5 & 175.4 \\
2.6 & 189.8 & 189.5 & 175.0 & 133.6 \\
3.0 & 118.0 & 116.4 & 106.1 & 80.4 \\
\enddata
\end{deluxetable}

\subsection{Masses, densities, and sizes}
\label{sec_masas}

\begin{figure}
\plotone{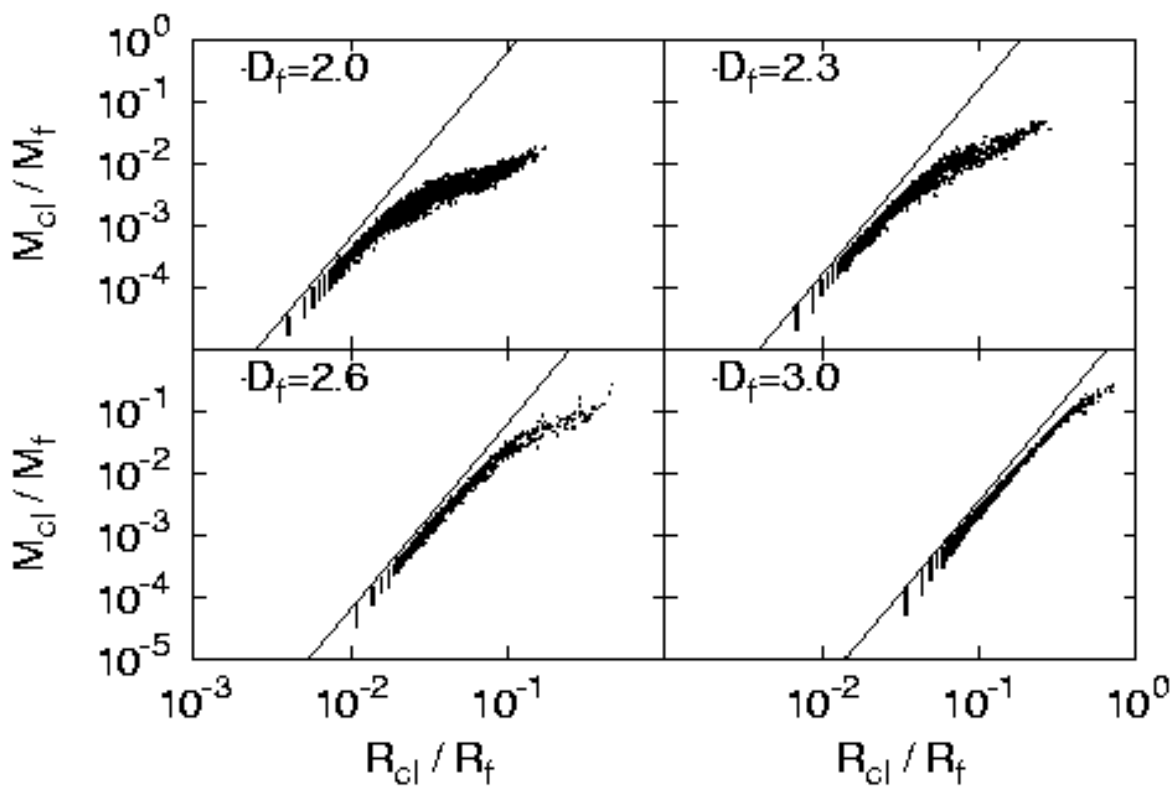}
\caption{Clump mass, $M_{cl}$, as a function of radius, $R_{cl}$,
         for the labeled values of fractal dimension $D_f$. The
         solid line corresponds to the constant density case.}
\label{fig_masa100}
\end{figure}

\begin{figure}
\plotone{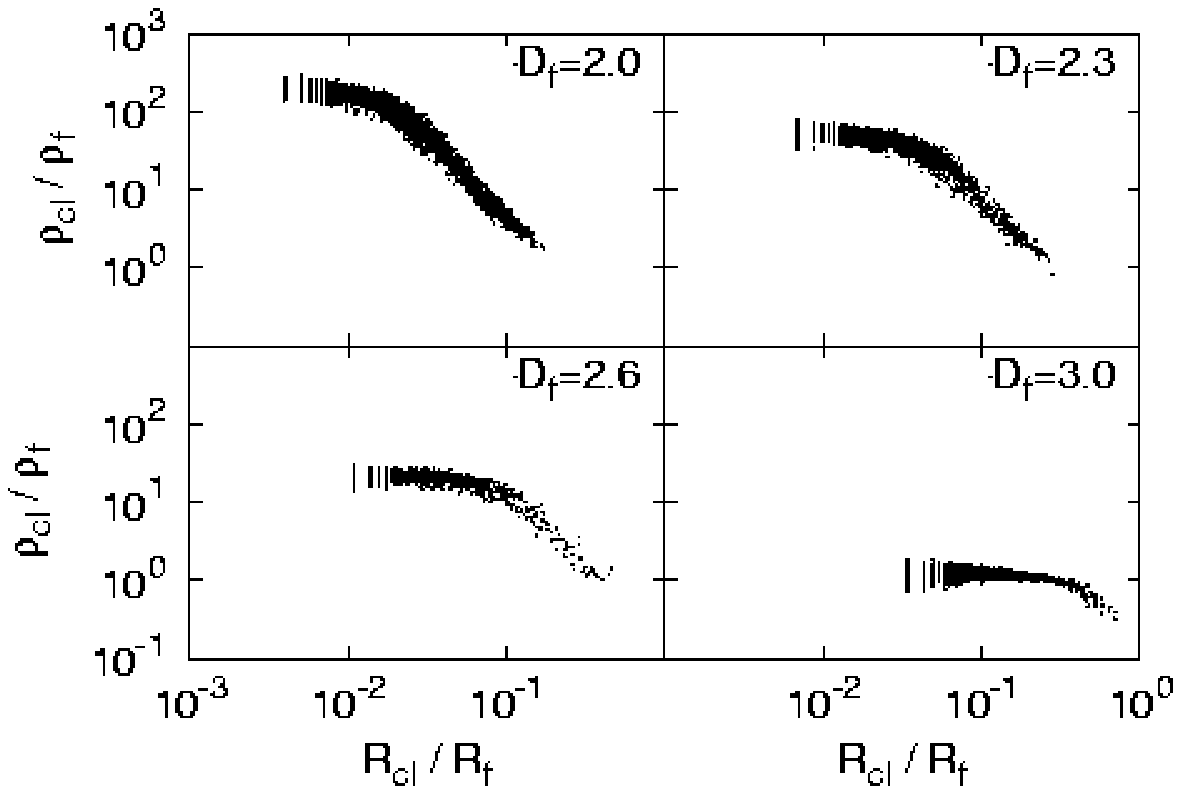}
\caption{The average clump density, $\rho_{cl}$, as a function of
         the radius, $R_{cl}$, for the labeled values of fractal
         dimension $D_f$.}
\label{fig_densi100}
\end{figure}

Generally the resulting clump shapes do not have a regular
geometry (see the examples in Figures~\ref{fig_mapa} and
\ref{fig_clumps100}). Under these circumstances it is not
easy to define a clump ``radius", so we have defined a
characteristic radius simply as the cubic root of the clump
volume. Figure~\ref{fig_masa100} shows
the mass of the clumps ($M_{cl}$) as a function of the radius
($R_{cl}$) for four different values of the fractal dimension
($D_f=2.0$, $2.3$, $2.6$, and $3.0$). The corresponding average
densities ($\rho_{cl}$) are shown in Figure~\ref{fig_densi100}.
The range of possible sizes depends in principle on the fractal
dimension (that determines the way in which fragmentation occurs) and
on the number of levels (that determines how many fragmentations occur).
When generating the fractal structure the smallest fragments have
radii $R_{min}/R_f = (1/N^H)^{1/D_f}$. Computational limitations
prevented us from using more than $H=9$ levels, so the smallest
structures have sizes $\sim 7 \times 10^{-3} R_f$ for $D_f=2.0$ and
$\sim 4 \times 10^{-2} R_f$ for $D_f=3.0$. The size range obtained
is more or less between these minimum values and the theoretical
maximum $R_{cl}/R_f=1$. Generally speaking, mass does not obey a
power law with size of the type $M_{cl} \varpropto R_{cl}^{\gamma}$
along all the size range (Figure~\ref{fig_masa100}), and therefore
neither density obeys the corresponding $\rho_{cl} \varpropto
R_{cl}^{\gamma - 3}$ (Figure~\ref{fig_densi100}). We would expect
$\gamma=D_f$ in the case of random sampling throughout the
hierarchy (section~\ref{sec_fractals}), but what we see is $\gamma
\simeq 3$ (i.e., $\rho \simeq constant$) at small sizes and then
$\gamma$ gradually decreases as $R_{cl}$ increases. For comparison,
we show in Figure~\ref{fig_masa100} the lines corresponding to 
$\rho_{cl} = constant$. The decrease in the $\gamma$ value is more
abrupt for low $D_f$ values, while for the extreme case $D_f=3$ we get
$\gamma \simeq 3$ along almost the full size range. For $D_f=3$ the
range in density values is very narrow and close to $\rho_{cl}/\rho_f
= 1$ (as expected), but for lower $D_f$ values the density range
increases and tends to higher values (Figure~\ref{fig_densi100}).
The latter situation corresponds to more fragmented structures with
small dense clumps separated by large low density regions.

An empirical power law scaling relation with
$\gamma \simeq 2$ was first pointed out by
\citet{lar81} for molecular cloud complexes (in the range
$0.1 \lesssim R \lesssim 100\ pc$ and $1 \lesssim M \lesssim
10^6\ M_{\sun}$), although it has been argued that this result
could be simply an artifact due to observational limitations
that strongly constraint the range of observed column densities
\citep{sca90}. This power law behavior has been generally
interpreted as a consequence of the mechanical equilibrium
in self-gravitating, turbulent molecular clouds, but also very
different physical processes can reproduce this type of
relations \citep[see][and references therein]{elm04}. From
the observational point of view, the problem is that many
studies show a weak or scatter-dominated correlation. Moreover,
comparing different observational studies is always difficult
because of the existing variety in observational techniques and
criteria used to define or derive properties such as mass, radius,
etc. \citet{elm96} used several surveys from the literature and
found $\gamma \simeq 2.35$, which was interpreted as a direct
consequence of an underlying fractal structure with $D_f \simeq
2.35$. However, this result only emerges when the clouds and
clumps from many regions are analyzed as an ensemble, while the
power laws measured in individual regions have $\gamma
\simeq 2.4-3.7$, with values $\gtrsim 3$ being typical.
\citet{rei05} found  $\gamma \simeq 1.5-2.1$ in the range
$1 \lesssim M \lesssim 10^3\ M_{\sun}$, whereas \citet{hei98}
found a steeper slope ($\gamma \simeq 2.3$) in the range
$10^{-4} \lesssim M \lesssim 10^2\ M_{\sun}$. \citet{cas95}
have already observed a change in the power law slope between
large mass and low mass cores in Orion A and B. \citet{fal04}
have argued that, in spite of the large scatter, it seems that
$\gamma \simeq 2$ better characterizes large scale structures
and $\gamma \simeq 2.3$ the small scale ones.

\begin{figure}
\plotone{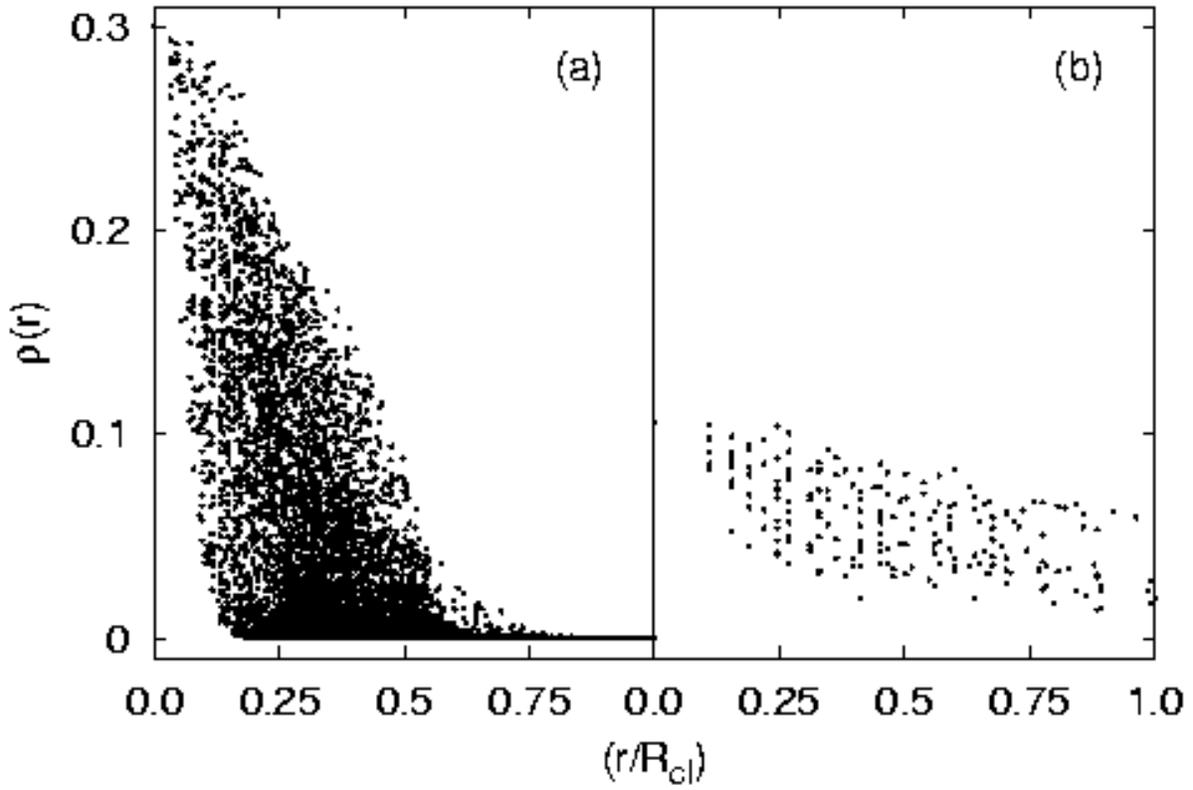}
\caption{The density of each pixel as a function of its distance
         (in units of the clump size)
         from the pixel having the maximum density value, for
         (a) a large and (b) a smaller clump in one of the
         fractals with $D_f=2.6$.}
\label{fig_perfiles}
\end{figure}

Our results seem to favor a change in the slope of the
$M_{cl}-R_{cl}$ relation, although less steep as $D_f$ increases.
The manner in which $M_{cl}$ increases with $R_{cl}$ is determined
by the density profile of the clumps. For flat density profiles,
$\rho (x,y,z) = constant$, we would expect to see $M_{cl} = 
\int_{V_{cl}} \rho (x,y,z) dV \sim R_{cl}^3$. For extremely
narrow density profiles of the form $\rho (x,y,z) = \rho_0 \delta
(x_0,y_0,z_0)$, being $\delta$ the delta of Dirac and $\rho_0$
the density value of a little region of volume $V_0$ around
$(x_0,y_0,z_0)$, we would have $M_{cl} = \int_{V_{cl}} \rho
(x,y,z) dV = \rho_0 V_0 = constant$. The observed behavior
is directly related to a gradual change in clump density
profiles from very flat shapes at small sizes to narrower
ones at large sizes. As an example, Figure~\ref{fig_perfiles}
shows density profiles for two different clumps in one of the
fractals with $D_f=2.6$. We have plotted the density of each
pixel as a function of its distance from the pixel having the
maximum density value. The scatter in these plots is due to
the asymmetry of the clump shapes. The figure labeled (a)
corresponds to a clump $\sim 40$ times more massive and 
$\sim 10$ times larger than the clump in (b), and the
characteristics mentioned before can readily be appreciated.
This is an interesting property of the fractal density
distribution we are generating, in which large clumps are the
result of the ``agglomeration" (through a gaussian convolution
with equation~\ref{ecu_kernel}) of particles in the last level
of hierarchy in the original fractal. Its consequences for
the derived clump properties will be discussed in
section~\ref{sec_threshold}.

\subsection{Clump mass and size spectra}

\begin{figure}
\plotone{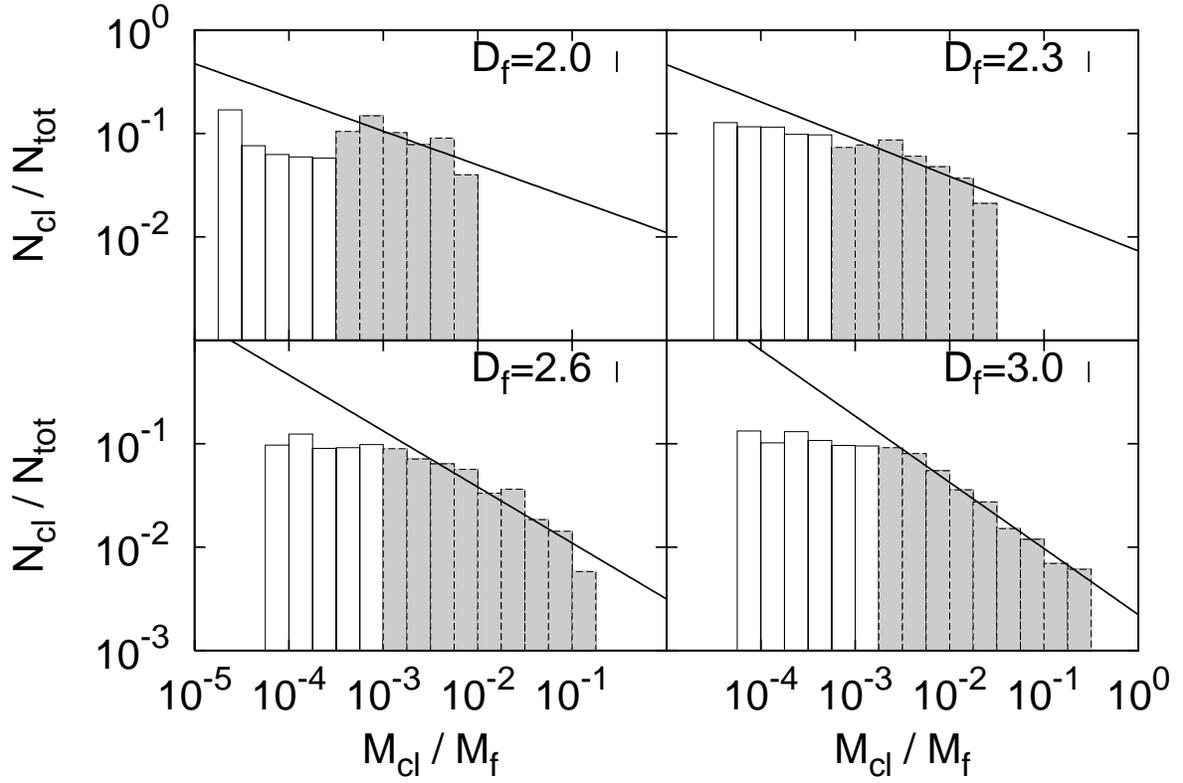}
\caption{The clump mass spectra for the labeled values of fractal
         dimension $D_f$. The filled histograms indicate
         the mass range for clumps having a volume larger than
         $\sim 10$ pixels,
         and the solid lines are the best fits in these ranges.}
\label{fig_dismas100}
\end{figure}

The clump mass spectra (number of clumps $N_{cl}$ per logarithmic
mass interval) for different $D_f$ values are shown in
Figure~\ref{fig_dismas100}.
The filled histograms indicate the ranges corresponding to
clumps having a volume larger than $\sim 10$ pixels, which were the
ranges where a power law-like behavior was observed:
$N_{cl} \varpropto M_{cl}^{-\alpha_M}$. The best fits in
these ranges (solid lines) show clearly steeper slopes as the
fractal dimension increases. The fits yielded:
$\alpha_M \simeq 0.33 \pm 0.12$ for $D_f=2.0$,
$\alpha_M \simeq 0.36 \pm 0.08$ for $D_f=2.3$,
$\alpha_M \simeq 0.54 \pm 0.06$ for $D_f=2.6$, and
$\alpha_M \simeq 0.64 \pm 0.03$ for $D_f=3.0$.
The mass spectra become flatter in the low mass range. We have to
point out that we are not saying that the {\it intrinsic} mass spectra
are actually power laws, but we are simply characterizing the spectra
with the functional form commonly used by observers in order to better 
compare these results.
From simple random sampling through the hierarchy we would expect
$\alpha_M = 1$ (see section~\ref{sec_fractals}),
but we already mentioned that if double counting is avoided
and/or random blending is taken into account the observed index
should be less than this value.
Clouds with high fractal dimensions are less fragmented than low
fractal dimension clouds. They have a relatively small number of
clumps but, on average, these clumps have larger sizes and higher
masses. As $D_f$ increases clouds tend to be more homogeneous, and
that is why the average clump densities are always close to the
whole cloud density (Figure~\ref{fig_densi100}). Additionally,
the volume in clumps approach the total cloud volume as $D_f$
increases (the filling factor tends to 1 as $D_f$ tends to 3),
and the clumps are relatively close to each other. On the opposite,
at very low fractal dimensions the high density clumps are separated
by low density (or empty at all) regions. Ultimately, low fractal 
dimension clouds distribute their material more homogeneously between
the larger number of well differentiated clumps. At high fractal
dimensions a greater amount of ``ambiguous" material is assigned
to massive clumps leaving a higher number of small clumps.
It is for this reason that the mass spectrum slope is flatter for
small $D_f$ values.

\begin{figure}
\plotone{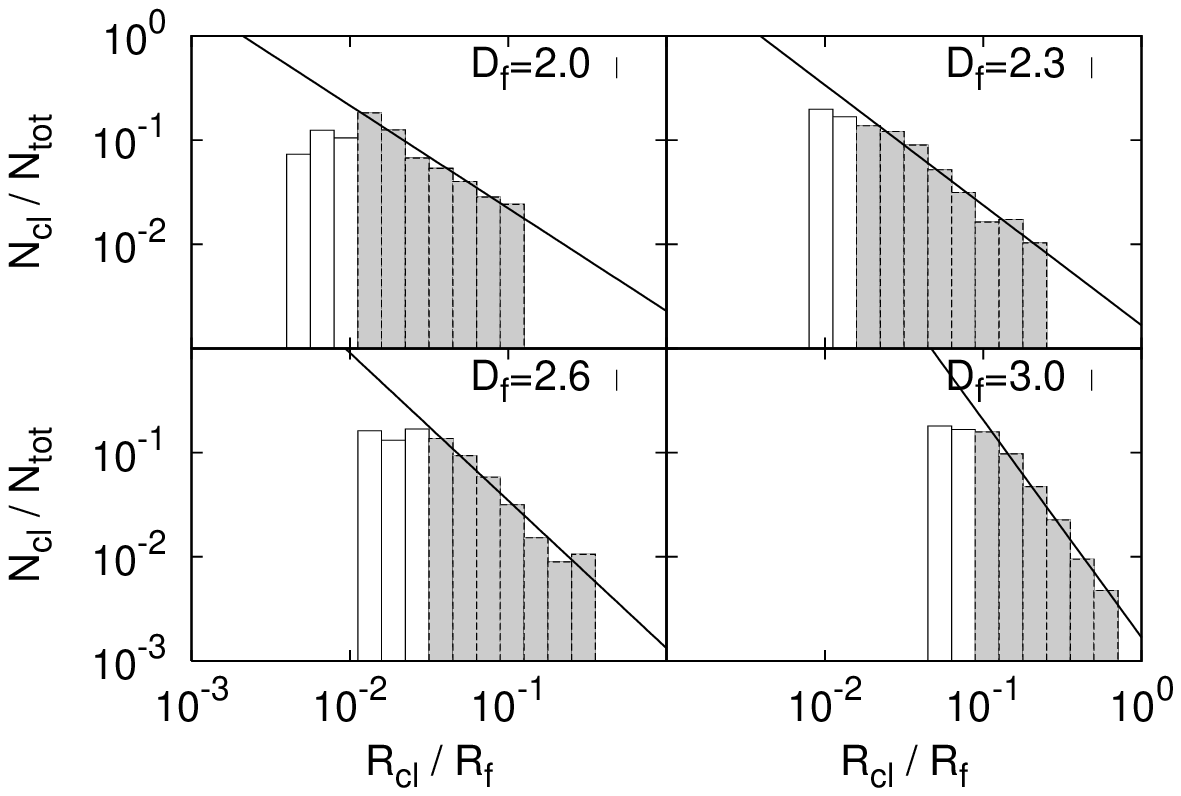}
\caption{The clump size spectra for the labeled values of fractal
         dimension $D_f$. The filled histograms indicate
         the size range for clumps having a volume larger than
         $\sim 10$ pixels, and the solid lines are the best fits
         in these ranges.}
\label{fig_disrad100}
\end{figure}

The clump size spectra (per logarithmic size interval) are shown
in Figure~\ref{fig_disrad100}
for the same fractal dimension values as in Figure~\ref{fig_dismas100}.
Again, the filled histograms indicate the ranges for clumps
larger than $\sim 10$ pixels, and the solid lines are the best fits
in these ranges. The near-power law behavior $N_{cl} \varpropto
R_{cl}^{-\alpha_R}$ becomes steeper as $D_f$ increases, as the
fitted slopes show:
$\alpha_R \simeq 0.99 \pm 0.07$ for $D_f=2.0$,
$\alpha_R \simeq 1.15 \pm 0.08$ for $D_f=2.3$,
$\alpha_R \simeq 1.42 \pm 0.13$ for $D_f=2.6$, and
$\alpha_R \simeq 2.09 \pm 0.08$ for $D_f=3.0$.
As for the mass spectra, the final distribution is flatter than
the random sampling slope which is $\alpha_R=D_f$
and, as before, the relative importance of the clump
selection criterion when applied to more or less fragmented
clouds causes the differences in the derived slopes for different
$D_f$ values. It is interesting to mention that if mass and
radius could be related as $M_{cl} \varpropto R_{cl}^{\gamma}$
then the power law indices satisfy a relationship of the form
$\alpha_R = \gamma \alpha_M$. For the random sampling case we
have $\gamma=D_f$, $\alpha_M=1$, and then $\alpha_R=D_f$, as
shown in section~\ref{sec_fractals}; but our results do not show
a scale relation between mass and radius ($\gamma \neq constant$,
see Figure~\ref{fig_masa100}) and therefore this kind of simple
relations cannot be found.

The power-law behavior obtained for both the mass and size
distributions are in general consistent with observations.
However, the detailed comparison is extremely
difficult considering the great diversity of results reported
in the literature and considering that no clear relationship has
been observed between the mass or size distributions and the mean
physical properties of the ISM. The most accepted view is a nearly
constant mass and size distributions throughout a large
range of scales, according to a fractal picture for the ISM
\citep[see, for instance, the reviews of][and references
therein]{eva99,wil00,fal04}. When a sufficiently large number
of clumps is considered, galactic surveys indicate that
on average $\alpha_M \sim 0.8$ in the range $1 \lesssim M
\lesssim 10^7\ M_{\sun}$ and $\alpha_R \sim 2.3$ in the range
$0.1 \lesssim R \lesssim 10^2\ pc$ \citep{elm96}, and these
results seem to depend very little, if any, on physical
properties like selfgravitation or the degree of star formation
activity \citep{sim01}. The extension at lower masses and sizes
yields $0.6 \lesssim \alpha_M \lesssim 0.8$ for $10^{-4}
\lesssim M \lesssim 10^3\ M_{\sun}$ and $\alpha_R \sim 2.0
\pm 0.2$ for $10^{-2} \lesssim R \lesssim 1\ pc$
\citep{hei98,kra98}. The mass and size distributions obtained 
here have indices in the range $0.3 \lesssim \alpha_M \lesssim
0.6$ and $1.0\lesssim \alpha_R \lesssim 2.1$ for $2 \leq D_f \leq
3$. If we assume $D_f \simeq 2.6$ then $\alpha_M \sim 0.5$ and 
$\alpha_R \sim 1.4$, a little lower but in gross agreement
with mean observed values. Steeper slopes can be obtained
only for $D_f > 2.6$. These results seem to favor the idea
that the ISM has a fractal dimension higher than the usually
assumed value $D_f \simeq 2.3$ \citep{san05}. However, the
situation is far from being totally understood. The mass 
distribution slopes may be as steep as, for example,
$\alpha_M \simeq 1.1$ for filamentary clumps in
Orion A molecular cloud \citep{nag98} or as flat as $\alpha_M
\simeq 0.3$ for the Rosette molecular cloud \citep{wil94}, not
being obvious whether these observations reveal real differences
from region to region. \citet{sch04} have demonstrated that the
derived properties differ significantly when using different
clump identification methods. The algorithm GAUSSCLUMPS
\citep{stu90} is able to separate blended clumps, although tends
to introduce spurious features if the structures do not have gaussian
shapes.  On the other hand, CLUMPFIND \citep{wil94} is able to
find arbitrarily shaped clumps but it does not detect less massive 
clumps which are blended with others. Thus, generally speaking, one
would expect steeper mass distributions for GAUSSCLUMPS than for
CLUMPFIND \citep{sch04}, but error bars are usually large and this
behavior cannot be appreciated \citep{moo04}. The situation
is worst for the size distributions because clump sizes are
quantities more difficult to estimate in a robust manner,
particularly for small, low intensity objects close to the
resolution limits \citep{elm96,hei98}.

\subsection{Dependence on the threshold density}
\label{sec_threshold}

The previous results refer to the case $\epsilon=1$. A smaller
$\epsilon$ value implies that not all the cloud complex mass has
formed clumps. We have simulated this effect by
considering a (constant) threshold density such that pixels
with densities below this threshold value are not taken into
account when constructing clumps. In principle, $\epsilon$
could be related to observational limitations, because smaller
$\epsilon$ values arise from the fact that there is a threshold
density level for observations. Also we could associate it with
the actual mass fraction in clumps, being the rest of the mass
kept in a diffuse interclump medium \citep{bli86,woo05}.
In such a case $\epsilon$ could be around $\sim 0.9$
or even much less \citep{bli86,pag98}. One possibility is to
assume that a fixed fraction of clump masses will form dense
prestellar cores and then $\epsilon \sim 0.1$ (if we assume
a star formation efficiency $\sim 10\%$), although
obviously the formation of gravitationally bounded cores
is a process more complex than this. The threshold densities
we discuss here have been chosen to obtain the following
$\epsilon$ values: $0.5$, $0.25$, and $0.1$, and all the
clump properties were recalculated.

\begin{figure}
\plotone{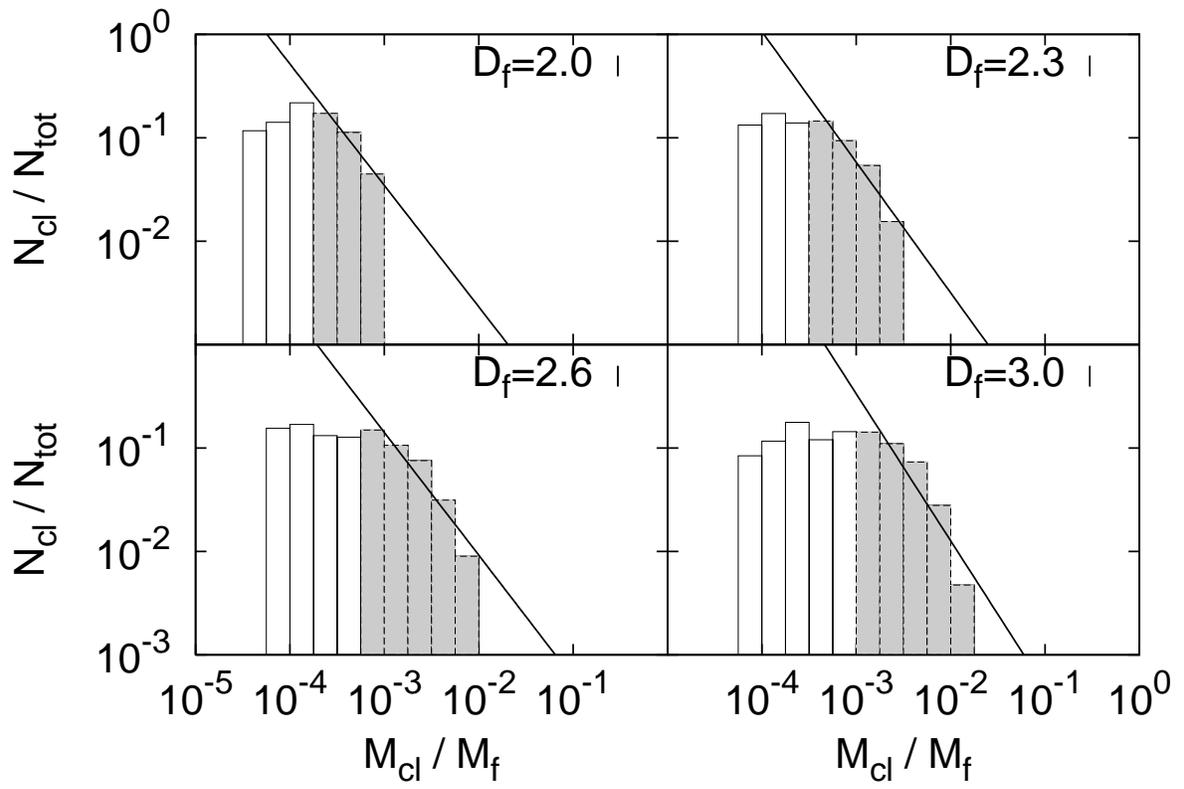}
\caption{The same as in Figure~\ref{fig_dismas100}, but for
         $\epsilon=0.1$.}
\label{fig_dismas010}
\end{figure}

\begin{figure}
\plotone{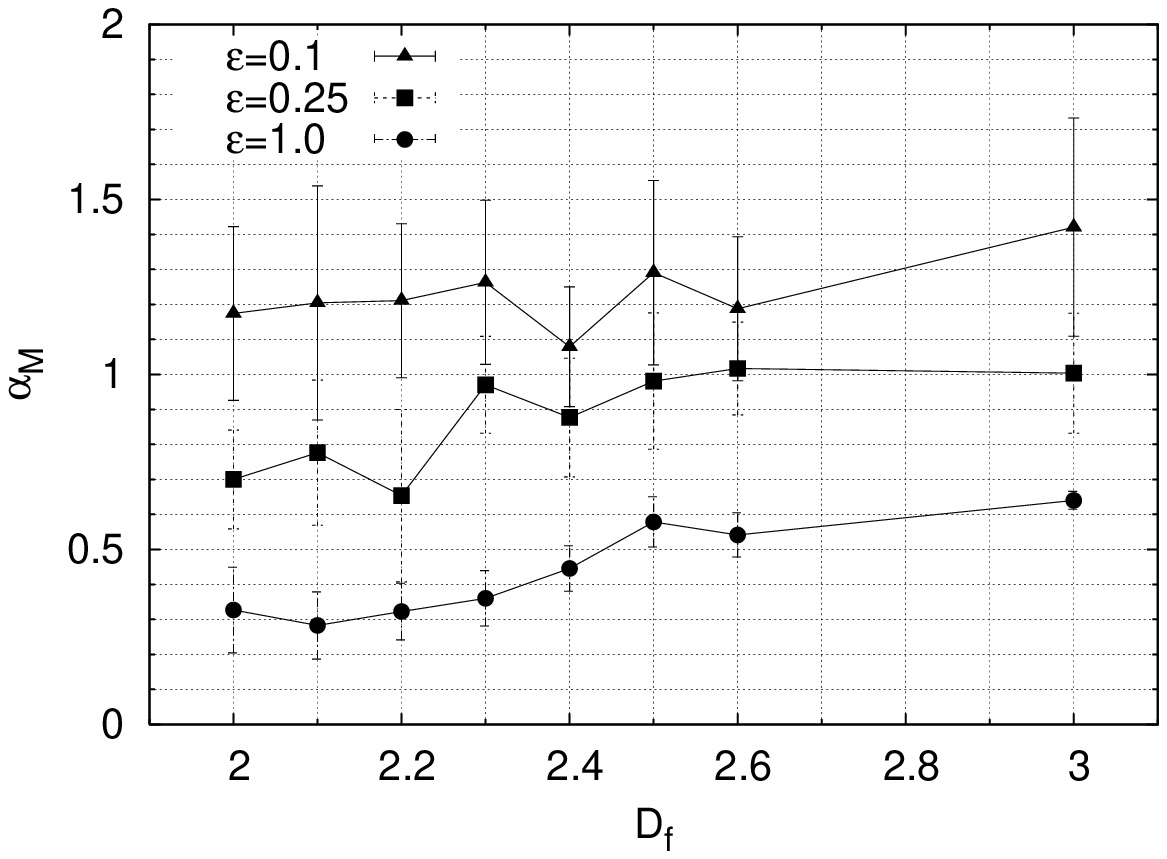}
\caption{The power law index $\alpha_M$ as a function of the fractal
         dimension $D_f$ for three different values of $\epsilon$:
         $1.0$ (circles), $0.25$ (squares), and $0.1$ (triangles).
         The bars on the data are the standard deviations of the
         best fits. The results for $\epsilon = 0.5$ are very close
         to those for $\epsilon = 1$ and are not shown for the sake
         of clarity.}
\label{fig_mass_slopes}
\end{figure}

For $\epsilon \leq 0.5$ the mass-radius relations for the
clumps correspond to the $\rho_{cl} = constant$ case.
The range of masses decreases, which is not a surprising
result since part of the clumps mass has been removed,
but additionally the mass spectra behavior modifies in an interesting
way: as $\epsilon$ decreases the slope in the power law range becomes
steeper. The mass spectra for the same fractal dimensions as before
but for $\epsilon=0.1$ are shown in Figure~\ref{fig_dismas010}.
The same general behavior is kept (a power law at high masses with a
flattening at low masses) but the slope values changed notoriously.
For example, for the case $D_f=2.6$ the index $\alpha_M$ increased
from $0.54 \pm 0.06$ ($\epsilon=1$) to $1.19 \pm 0.20$ ($\epsilon=
0.1$). We have summarized all the obtained mass spectra by plotting
in Figure~\ref{fig_mass_slopes}
the power law slope $\alpha_M$ as a function of the fractal dimension
$D_f$ for three different values of $\epsilon$ ($1.0$, $0.25$, and
$0.1$). The result for $\epsilon=0.5$ is so close to the result for
$\epsilon=1$ that it is not shown for simplicity. As already mentioned,
the general behavior is an increase of $\alpha_M$ (steeper slopes) as
$D_f$ increases, but also $\alpha_M$ increases as $\epsilon$ decreases.
In spite of the bar sizes, Figure~\ref{fig_mass_slopes} allows us
to quantify approximately the dependence of $\alpha_M$ on both $D_f$
and $\epsilon$. Part of the uncertainty comes from fitting power law 
functions to distributions that depart from this behavior.

\begin{figure}
\plotone{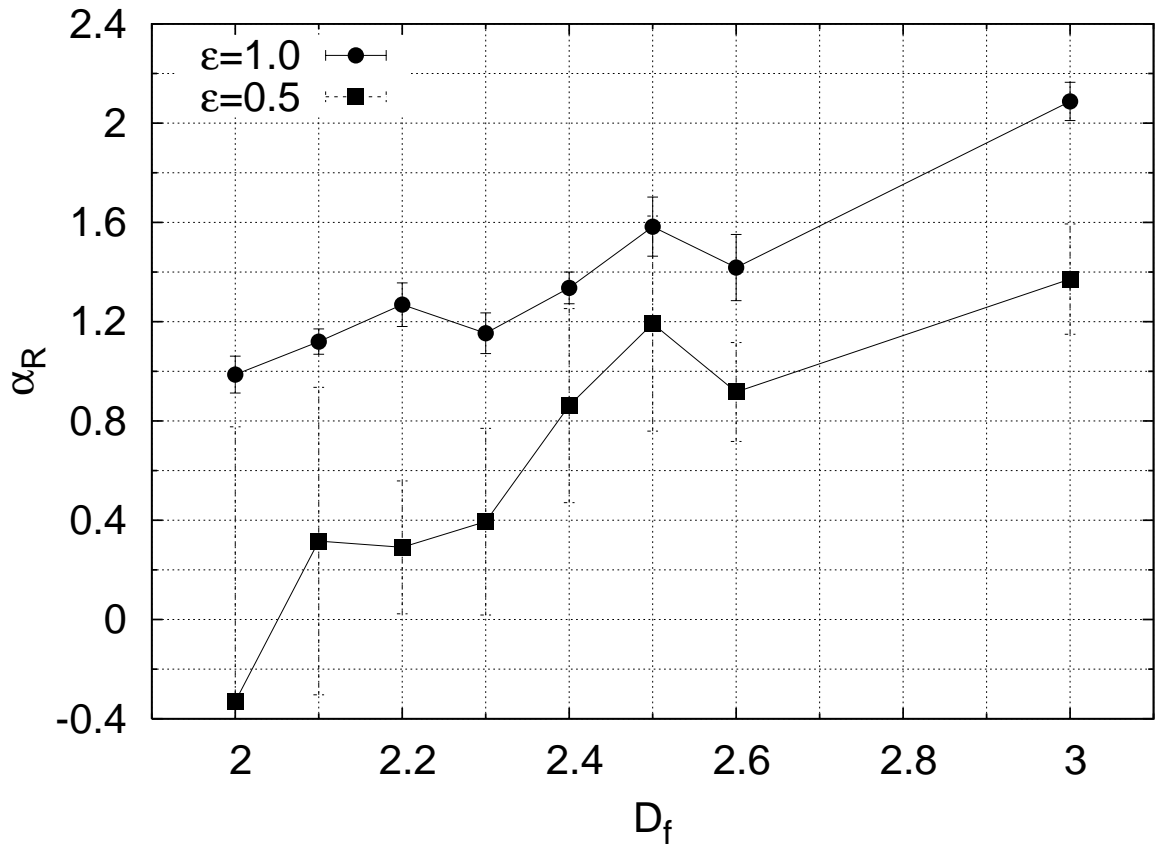}
\caption{The power law index $\alpha_R$ as a function of the fractal
         dimension $D_f$ for two different values of $\epsilon$:
         $1.0$ (circles), and $0.5$ (squares). The bars on the data
         are the standard deviations of the best fits.}
\label{fig_size_slopes}
\end{figure}

Concerning the clump size spectra, when $\epsilon=0.5$ the size
range decreases in such a way that a clear power law behavior is
not seen, mainly at low fractal dimensions. However, we
have followed the objective procedure of fitting a power law for
structures bigger than $\sim 10$ pixels, and we obtained flatter 
distributions. For example, for $D_f=2.6$ we have that $\alpha_R$
decreased from $1.42 \pm 0.13$ ($\epsilon=1$) to $0.92 \pm 0.20$ 
($\epsilon=0.5$). The obtained power law slopes $\alpha_R$ as a function
of the fractal dimensions $D_f$ for $\epsilon = 1$ and $\epsilon = 0.5$
are shown in Figure~\ref{fig_size_slopes}.
For smaller $\epsilon$ values the size ranges were so narrow that
fits could not be done and/or standard deviation bars were extremely
large. The power law index $\alpha_R$ increases as $D_f$ increases, but
now we have that $\alpha_R$ also decreases when $\epsilon$ decreases,
in opposite to the mass spectra.

The trend of steeper mass distributions for higher threshold densities
(smaller $\epsilon$ values) was seen by \citet{elm02} in his model of
fractal brownian motion clouds, although he only simulated clouds with
constant fractal dimension ($D_f \simeq 2.75$ on average). It is not
easy to understand the dependence of $\alpha_M$ and $\alpha_R$ on
$\epsilon$, but as in the work of \citet{elm02} it has to do with
changes in the density profiles between the smallest and largest
clumps. In section~\ref{sec_masas} we argued that density profiles
of small clumps are flatter than those in large, massive clumps
(which have lesser average density). Large clumps have narrow,
tall profiles with long tails at low densities (see the example
in Figure~\ref{fig_perfiles}). Then, when the threshold density is
increased, both sizes and masses decrease, but in different ways.
Large clumps decrease their sizes in a higher proportion than the
small ones, because of the long tails. Thus, the size distribution 
flattens because the proportion of high radius clumps increases
(relative to the small ones). On the opposite, there is not too 
much mass below the low density tails and then the masses of large
clumps decrease in a lower proportion than for smaller clumps.
Therefore, when the threshold increases the proportion of high
mass clumps decreases and the mass distribution becomes steeper.

We have seen that
for more realistic $\epsilon$ values (i.e., $\epsilon < 1$) we get
steeper mass functions and flatter size functions. The case $\epsilon
=0.5$ and $D_f=2.6$ yields $0.7 \lesssim \alpha_R \lesssim 1.1$,
which is much flatter than the average observed in molecular clouds
but remarkably similar to the size distribution recently observed
for HI clouds in the LMC \citep{kim05}. Additionally, according to
our results, molecular cloud complexes with $D_f=2.6$ will have
mass functions $\alpha_M \sim 1.2 \pm 0.2$ when $\epsilon = 0.1$,
very close to the value $1.35$ of \citet{sal55}. In fact, the
observed mass spectrum of dense cores seems to have a form
significantly different from molecular clumps, with the slope 
relatively flatter ($\alpha_M \simeq 0.5$) at low masses and steeper
($\alpha_M \simeq 1.5$) at high masses \citep{mot98,rei05}. Our
results reproduce this behavior, we clearly appreciate (in spite
of the large standard deviation bars) a systematically steeper mass
function at high masses as $\epsilon$ decreases. This is a very 
conspicuous result: the denser regions in the fractal cloud complex 
have a Salpeter-like mass function steeper than the mass function of
the whole complex. An interesting point is that the mass distribution
slope for $\epsilon=0.1$ is near to the Salpeter value (within the
error bars) in a wide range of fractal dimension values
($2.0 \lesssim D_f \lesssim 2.6$). It seems that the information
about the fractal density structure is lost when we only see the
central dense ``cores". The results we showed in this section 
suggest a direct link between cloud structure and star formation.

\section{CONCLUSIONS}
\label{sec_conclusion}

In this work we have studied the relation between the physical
properties of interstellar cloud complexes and their underlying
fractal structure. For a better understanding of this relationship
we have used a simple algorithm that generates fractal clouds with
fractal dimensions well defined in a wide range of space scales.
We have not considered physical processes to generate either the
fractal structure or the embedded clumps. In any case, the exact
physical nature of clumps still remains uncertain; whether they
are temporary density fluctuations caused by supersonic turbulence
or more stable structures confined by the interclump medium
\citep{wil00}. The simple approach given here allows us to analyze
in an empirical way the dependence of the ISM properties on both
the fractal dimension and the threshold density.

We observe that the number of clumps (as given by the number of
relative maxima of density) depends only on the density structure
of the whole cloud, which can be associated to a single parameter:
the fractal dimension. However, when the whole mass of the cloud
is distributed amongst the different clumps (i.e., $\epsilon=1$),
the mass and size distributions are highly dependent on the clump
defining criterium chosen. On the other hand, as the fraction of
the total mass in the form of clumps ($\epsilon$) decreases the
mass and size distributions become similar for different criteria
of clump selection, i.e., when only the ``cores" of the ``clumps"
are selected their distributions in mass and radius are not 
dependent on the selection criteria. This fact leads to an
interesting conclusion: the ``cores" mass and size distributions
are only driven by the fractal dimension of the clouds. A true
molecular cloud contains relatively empty voids, dense cores,
as well as rarefied intercloud material \citep{gam03}, which looks
like more similar to the distributions of cores obtained for
$\epsilon=0.1$ than the clumps distribution for $\epsilon=1$.

In general, the masses and radii of the resulting clumps do not
fulfill simple power law relations of the type $M_{cl} \varpropto
R_{cl}^{\gamma}$ along all the size range. We obtain that $\gamma 
\simeq 3$ at small sizes and $\gamma < 3$ at larger sizes with the
exact behavior depending on the fractal dimension. The reason for
the departure from a power law has to do with the differences in
the density profiles between the smallest and largest clumps: small
clumps have flat density profiles whereas large clumps have narrow
and tall profiles with long tails at low densities.

The number of clumps per logarithmic interval of mass (or size)
can be fitted to a power law function $N_{cl} \varpropto
M_{cl}^{-\alpha_M}$ (or $N_{cl} \varpropto R_{cl}^{-\alpha_R}$)
in a certain range of mass (or size). The indices $\alpha_M$
and $\alpha_R$ depend both on the fractal dimension ($D_f$) and on
the fraction of the total mass in the form of clumps ($\epsilon$).
Our main results are summarized in Figures~\ref{fig_mass_slopes}
and \ref{fig_size_slopes}, and they refer to the dependence of
the clump mass and size spectra on both $D_f$ and $\epsilon$.
For the case $\epsilon=1$ we obtain that as $D_f$
increases from $2$ to $3$ $\alpha_M$ increases from $\sim 0.3$ to
$\sim 0.6$ whereas $\alpha_R$ increases from $\sim 1.0$ to $\sim 2.1$. 
Rough comparison with observations suggests that $D_f \simeq 2.6$ is
consistent with the average properties of the ISM. This value
for the ISM fractal dimension is in agreement with previous
results based on the fractal dimension measured on the
projected images of clouds \citep{san05}. On the other hand, as
$\epsilon$ decreases $\alpha_M$ increases and $\alpha_R$ decreases.
For the case $\epsilon = 0.1$ (only $10\%$ of the complex mass is
in the form of dense clumps) we obtain $\alpha_M \simeq 1.2$ for
$D_f=2.6$, a value remarkably similar to the \citet{sal55} value,
suggesting that the stellar initial mass function could be
intimately related to the internal structure of molecular cloud
complexes.

In summary, we have derived the properties of fractal cloud
complexes, no mattering the physical processes generating the
fractal structure. It seems that $D_f \simeq 2.6$ is consistent
with observations, but the relevance of this relatively high
fractal dimension value has to be analyzed in future studies,
mainly concerning the physical processes involved in the
structure of the ISM.

\acknowledgments
We want to thank the referee for his/her helpful comments and
criticisms which greatly improved this paper.
N.~S. would like to acknowledge the funding provided by the
Secretar\'{\i}a de Estado de Universidades e Investigaci\'on
(Spain) through grant SB-2003-0239.
E.~J.~A. acknowledges the funding from MECyD of Spain through
grants AYA2004-05395 and AYA2004-08260-C03-02, and from
Consejer\'{\i}a de Educaci\'on y Ciencia (Junta de Andaluc\'{\i}a)
through TIC-101.
E.~P. acknowledges financial support from grants
AYA2004-02703 and TIC-114.


\begin{thebibliography}{}

\bibitem[Ballesteros-Paredes(2004)]{bal04}
         Ballesteros-Paredes, J. 2004,
         \apss, 292, 193
\bibitem[Beech(1992)]{bee92}
         Beech, M. 1992,
         \apss, 192, 103
\bibitem[Bensch et al.(2001)]{ben01}
         Bensch, F., Stutzki, J., \& Ossenkopf, V. 2001,
         \aap, 366, 636
\bibitem[Beresnyak et al.(2005)]{ber05}
         Beresnyak, A., Lazarian, A., \& Cho, J. 2005,
         \apj, 624, L93
\bibitem[Blitz \& Stark(1986)]{bli86}
         Blitz, L., \& Stark, A. A. 1986,
         \apj, 300, L89
\bibitem[Blitz \& Williams(1997)]{bli97}
         Blitz, L., \& Williams, J. P. 1997,
         \apj, 488, L145
\bibitem[Brunt \& Heyer(2002a)]{bru02a}
         Brunt, C. M., \& Heyer, M. H. 2002a,
         \apj, 566, 276
\bibitem[Brunt \& Heyer(2002b)]{bru02b}
         Brunt, C. M., \& Heyer, M. H. 2002b,
         \apj, 566, 289
\bibitem[Caselli \& Myers(1995)]{cas95}
         Caselli, P., \& Myers, P. C. 1995,
         \apj, 446, 665
\bibitem[de Vega et al.(1996)]{dev96}
         de Vega, H. J., Sanchez, N., \& Combes, F. 1996,
         \nat, 383, 56
\bibitem[Elmegreen(1997a)]{elm97a}
         Elmegreen, B. G. 1997a,
         \apj, 477, 196
\bibitem[Elmegreen(1997b)]{elm97b}
         Elmegreen, B. G. 1997b, 
         \apj, 486, 944
\bibitem[Elmegreen(1999)]{elm99}
         Elmegreen, B. G. 1999,
         \apj, 515, 323
\bibitem[Elmegreen(2002)]{elm02}
         Elmegreen, B. G. 2002,
         \apj, 564, 773
\bibitem[Elmegreen \& Falgarone(1996)]{elm96}
         Elmegreen, B. G., \& Falgarone, E. 1996,
         \apj, 471, 816
\bibitem[Elmegreen \& Scalo(2004)]{elm04}
         Elmegreen, B. G., \& Scalo, J. 2004,
         \araa, 42, 211
\bibitem[Evans(1999)]{eva99}
         Evans, N. J. II 1999,
         \araa, 37, 311
\bibitem[Falgarone et al.(2004)]{fal04}
         Falgarone, E., Hily-Blant, P., \& Levrier, F. 2004,
         \apss, 292, 89
\bibitem[Falgarone et al.(1991)]{fal91}
         Falgarone, E., Phillips, T. G., \& Walker, C. K. 1991,
         \apj, 378, 186
\bibitem[Fischera \& Dopita(2004)]{fis04}
         Fischera, J., \& Dopita, M. A. 2004,
         \apj, 611, 919
\bibitem[Fleck(1996)]{fle96}
         Fleck, R. C. 1996,
         \apj, 458, 739
\bibitem[Gammie et al.(2003)]{gam03}
         Gammie, C. F., Lin, Y.-T., Stone, J. M., \& Ostriker, E. C.
         \apj, 592, 203
\bibitem[Heiles \& Troland(2005)]{hei05}
         Heiles, C., \& Troland, T. H. 2005,
         \apj, 624, 773
\bibitem[Heithausen et al.(1998)]{hei98}
         Heithausen, A., Bensch, F., Stutzki, J., Falgarone, E., \&
         Panis, J. F. 1998,
         \aap, 331, L65
\bibitem[Henriksen(1986)]{hen86}
         Henriksen, R. N. 1986,
         \apj, 310, 189
\bibitem[Henriksen(1991)]{hen91}
         Henriksen, R. N. 1991,
         \apj, 377, 500
\bibitem[Hentschel \& Procaccia(1982)]{hen82}
         Hentschel, H. G. E., \& Procaccia, I. 1982,
         \prl, 49, 1158
\bibitem[Kim et al.(2005)]{kim05}
         Kim, S., Staveley-Smith, L., Dopita, M. A., Sault, R. J.,
         Freeman, K. C., Lee, Youngung, \& Chu. Y.-H. 2005,
         preprint (astro-ph/0506224)
\bibitem[Kramer et al.(1998)]{kra98}
         Kramer, C., Stutzki, J., Rohrig, R., \& Corneliussen, U. 1998,
         \aap 329, 249
\bibitem[Larson(1981)]{lar81}
         Larson, R. B. 1981,
         \mnras, 194, 809
\bibitem[Larson(2003)]{lar03}
         Larson, R. B. 2003,
         Rep. Prog. Phys., 66, 1651
\bibitem[Lee(2004)]{lee04}
         Lee, Y. 2004,
         JKAS, 37, 137
\bibitem[Li et al.(2003)]{li03}
         Li, Y., Klessen, R. S., \& Mac Low, M.-M, 2003,
         \apj, 592, 975
\bibitem[Meneveau \& Sreenivasan(1990)]{men90}
         Meneveau, C.,\& Sreenivasan, K. R. 1990,
         \pra, 41, 2246
\bibitem[Mookerjea et al.(2004)]{moo04}
         Mookerjea, B., Kramer, C., Nielbock, M., Nyman, L.-A. 2004,
         \aap, 426, 119
\bibitem[Motte et al.(1998)]{mot98}
         Motte, F., Andre, P., \& Neri, R. 1998,
         \aap, 336, 150
\bibitem[Nagahama et al.(1998)]{nag98}
         Nagahama, T., Mizuno, A., Ogawa, H., \& Fukui, Y. 1998,
         \apj, 116, 336
\bibitem[Ostriker et al.(2001)]{ost01}
         Ostriker, E. C., Stone, J. M., \& Gammie, C. F. 2001,
         \apj, 546, 980
\bibitem[Padoan et al.(2004)]{pad04}
         Padoan, P., Jimenez, R., Nordlund, A., \& Boldyrev, S. 2004,
         \prl, 92, 191102
\bibitem[Padoan \& Nordlund(2002)]{pad02}
         Padoan, P., \& Nordlund, A. 2002,
         \apj, 576, 870
\bibitem[Pagani(1998)]{pag98}
         Pagani, L. 1998,
         \aap, 333, 269
\bibitem[Passot \& Vazquez-Semadeni(1998)]{pas98}
         Passot, T. \& Vazquez-Semadeni, E. 1998,
         \pre, 58, 4501
\bibitem[Passot \& Vazquez-Semadeni(2003)]{pas03}
         Passot, T. \& Vazquez-Semadeni, E. 2003,
         \aap, 398, 845
\bibitem[Reid \& Wilson(2005)]{rei05}
         Reid, M. A., \& Wilson, C. D. 2005,
         \apj, 625, 891
\bibitem[Salpeter(1955)]{sal55}
         Salpeter, E. E. 1955,
         \apj, 121, 161
\bibitem[S\'anchez et al.(2005)]{san05}
         S\'anchez, N., Alfaro, E. J., \& P\'erez, E. 2005,
         \apj, 625, 849
\bibitem[Scalo(1990)]{sca90}
         Scalo, J. 1990,
         in Physical Processes in Fragmentation and Star Formation,
         ed. R. Capuzzo-Dolcetta, C. Chiosi \& A. Di Fazio
         (Dordrecht: Kluwer), 151
\bibitem[Scalo et al.(1998)]{sca98}
         Scalo, J., Vazquez-Semadeni, E., Chappell, D., \&
         Passot, T. 1998,
         \apj, 504, 835
\bibitem[Schneider \& Brooks(2004)]{sch04}
         Schneider, N., \& Brooks, K. 2004,
         PASA, 21, 290
\bibitem[Silvermann(1986)]{sil86}
         Silvermann B. W. 1986, Density Estimation for Statistics and
         Data Analysis (New York: Chapman \& Hall)
\bibitem[Simon et al.(2001)]{sim01}
         Simon, R., Jackson, J. M., Clemens, D. P., Bania, T. M., \&
         Heyer, M. H. 2001,
         \apj, 551, 747
\bibitem[Slyz et al.(2005)]{sly05}
         Slyz, A. D., Devriendt, J. E. G., Bryan, G., \&
         Silk, J. 2005,
         \mnras, 356, 737
\bibitem[Stutzki et al.(1998)]{stu98}
         Stutzki, J., Bensch, F., Heithausen, A., Ossenkopf, V.,
         \& Zielinsky, M. 1998,
         \aap, 336, 697 
\bibitem[Stutzki \& Gusten(1990)]{stu90}
         Stutzki, J., \& Gusten, R. 1990,
         \apj, 356, 513
\bibitem[Vazquez-Semadeni(1994)]{vaz94}
         Vazquez-Semadeni, E. 1994,
         \apj, 423, 681
\bibitem[Wada \& Norman(2001)]{wad01}
         Wada, K., \& Norman, C. A. 2001,
         \apj, 547, 172
\bibitem[Williams et al.(2000)]{wil00}
         Williams, J. P., Blitz, L., \& McKee, C. F. 2000,
         in Protostars and Planets IV,
         ed. V. Mannings, A. P. Boss \& S. S. Russell
         (Tucson: University of Arizona Press), 97
\bibitem[Williams et al.(1994)]{wil94}
         Williams, J. P., de Geus, E. J., \& Blitz, L. 1994,
         \apj, 428, 693
\bibitem[Wood et al.(2005)]{woo05}
         Wood, K., Haffner, L. M., Reynolds, R. J., Mathis, J. S., \&
         Madsen, G. 2005,
         \apj, 633, 259

\end{thebibliography}
\end{document}